\documentclass[aps,prb,twocolumn,groupedaddress]{revtex4-1}

\usepackage[utf8]{inputenc}
\usepackage{amsmath}
\usepackage{units}
\usepackage{graphicx}

\renewcommand{\i}{\mathrm{i}}	
\newcommand{\e}{\mathrm{e}}		
\newcommand{\abs}[1]{\ensuremath{\left\vert#1\right\vert}}
\newcommand{\chem}[1]{\ensuremath{\mathrm{#1}}}
\newcommand{\un}[1]{\unit{#1}}
\newcommand{\ab}[1]{\mathrm{#1}}
\newcommand{\vect}[1]{\mathbf{#1}}
\newcommand{\upd}{\mathrm{d}}
\begin{document}
\title{Measuring the plasma-wall charge by infrared spectroscopy}

\author{K. Rasek}
\author{F. X. Bronold}
\affiliation{Institut f{\"u}r Physik,
	Universit{\"a}t Greifswald, 17487 Greifswald, Germany }
\author{M. Bauer}
\affiliation{Institut f\"ur Experimentelle und Angewandte Physik, Christian-Albrechts-Universit\"at zu Kiel - 24098 Kiel, Germany}
\author{H. Fehske}
\affiliation{Institut f{\"u}r Physik,
	Universit{\"a}t Greifswald, 17487 Greifswald, Germany }
\date{\today}

\begin{abstract}
We show that the charge accumulated by a dielectric plasma-facing solid can be measured 
by infrared spectroscopy. The approach utilizes a stack of materials 
supporting a surface plasmon resonance in the infrared. For frequencies near the Berreman 
resonance of the layer facing the plasma the reflectivity dip--measured from the back of the
stack, not in contact with the plasma--depends strongly on the angle of incidence making
it an ideal sensor for the changes of the layer's dielectric function due to the 
polarizability of the trapped surplus charges. The charge-induced shifts of the dip, 
both as a function of the angle and the frequency of the incident infrared light, are 
large enough to be measurable by attenuated total reflection setups.
\end{abstract}

\maketitle

\section{Introduction}
Fundamental to any interface is charge separation. This universal mantra
holds also for solids facing an ionized gas where an electron-depleted region in 
front of the solid is balanced by an electron-rich region inside or on top of the 
solid depending on the solid's electronic structure. The electron-depleted, positive 
part of the double layer--the plasma sheath--has been studied rather extensively in the 
past, in particular, its merging with the bulk plasma~\cite{Brinkmann09,Franklin03,Riemann91}.
But the negative part--the wall charge--and its merging with the bulk of the solid received little 
attention~\cite{BF17}, although it is an integral part of the electric response of the 
plasma-solid interface and thus unavoidably linked to the overall charge balance of the 
discharge. Especially the behavior of microdischarges integrated on semiconducting 
substrates~\cite{EPC13,DOL10} may be strongly affected by the charge dynamics inside the 
substrate. However, to develop an understanding of it requires experimental techniques 
probing inside the solid. So far only a few attempts have been made to measure the charge
accumulated by a solid in contact with a plasma. Besides traditional electric probes~\cite{KA80} 
and micron-size opto-mechanical charge sensors~\cite{PFS96}, which both utilize the principle 
of electric influence, the opto-electric Pockels effect~\cite{KTZ94} has been used for that purpose. 
The latter was developed into a rather sophisticated tool for lateral imaging of the 
wall charge in barrier discharges~\cite{TBW14}. It works however only for dielectric
coatings featuring the Pockels effect. For the dielectrics typically used in 
low-temperature plasma physics--\chem{SiO_2} and \chem{Al_2O_3}--it is not 
applicable. The semiconductors hosting the arrays of microdischarges referred to above
are also not Pockels-active. 

In this work we propose an infrared diagnostics for the charge collected by plasma-facing 
dielectrics which also works for the standard materials used in plasma physics. It 
utilizes the charge-sensitivity of the infrared reflectivity of a layered 
structure in contact with a plasma, where the plasma-facing, charge-collecting layer is 
made out of the dielectric of interest. Its width is chosen such that it supports a 
Berreman mode~\cite{Berreman}, thereby making the device sensitive to the low charge 
densities expected at plasma-solid interfaces compared to the rather high densities at 
solid-electrolyte interfaces~\cite{Chazalviel01,GE80} and semiconductor surfaces~\cite{WP12,JMS16},
to which such an arrangement could be also applied. Using an attenuated total reflection (ATR) 
spectroscopy setup enables us to utilize as a charge diagnostics not only the charge-sensitive 
frequency shift of the Berreman mode but also the shift of the angle of incidence where the 
mode occurs. 

\begin{figure}
	\centering
	\includegraphics{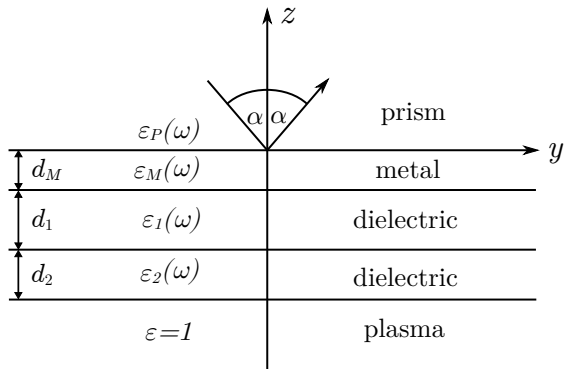}
	\caption{Structural composition of the system under consideration. The prism and metal
	layer allow plasmon resonance. The dielectric materials are chosen so that the surplus 
	charges are confined to the plasma-facing layer, which is the material of interest.}
	\label{fig.1}
\end{figure}

The stack of materials comprising the measuring device, which we envisage to be 
inserted into the plasma wall or the electrode, is shown in fig.~\ref{fig.1}. Due to the 
metal layer and the optical prism on top of it surface plasmon polaritons (SPPs) are excited 
which--by avoided resonance crossing with the Berreman mode of the layer facing the plasma
and consisting of the material of interest--cause 
a strong dependence of the reflectivity of the stack on the angle of incidence. If one is 
interested only in measuring the total charge collected by the dielectric the charge 
can be confined to a narrow region using a rather thin plasma-facing layer separated 
from the metal by a dielectric with negative electron affinity producing thereby a potential 
well. In case the density profile normal to the interface should also be mapped out, 
the plasma-facing layer has to be thick enough to host the profile yet thin enough to still 
support the Berreman resonance. 

To demonstrate the feasibility of the proposal we calculate the reflectivity of the 
structure shown in fig.~\ref{fig.1} as a function of the angle of incidence and the 
wavenumber (which we use in the final plots instead of the frequency) 
of the incident electromagnetic wave assuming--for simplicity--the surplus charges  
distributed homogeneously in the plasma-facing layer. The surplus charges' polarizability,
which we obtain from a memory function approach taking electron-phonon scattering into 
account~\cite{Memory}, modifies the dielectric function of the layer and is the ultimate 
reason for the charge-induced shifts of the Berreman mode. In the infrared the shifts we obtain 
are large enough to be detectable with common reflectivity setups. For the proof of principle
presented in this work we focus on measuring the total charge and not the whole density profile. 
It would require a more sophisticated theoretical treatment, taking nonlocal surface effects
of the electromagnetic response of the charged stack into account,
and is left for future work. The widths of the layers can thus 
be used almost freely as parameters to optimize the sensitivity of the setup.

\section{Theoretical description}
The physical process enabling the structure shown in fig.~\ref{fig.1} to be used as 
a charge sensing device is the interaction of the surface plasmon resonance (SPR) of 
the metallic layer below the prism and the Berreman mode of the plasma-facing,
charge-carrying dielectric layer. To identify suitable materials to be stacked together 
we start the description of our proposal with a discussion of the role of each layer. 
The prism and the metallic layer are essential for the SPR. They constitute a 
Kretschmann configuration~\cite{Kretschmann}, where total reflection of the 
incident wave at the prism-metal interface creates the evanescent wave necessary 
for exciting a SPP at the metal-dielectric interface~\cite{Kooyman08}. For SPR 
the wave extending into the plasma needs to be evanescent as well. Hence, the total 
reflection condition $\sin^2\alpha > 1/\varepsilon_\mathrm{P}$ imposes a lower limit 
to the angle of incidence, depending on the prism material. Because of this relation,
the dielectric function of the prism should be nearly independent of frequency 
$\omega$ in the range of interest. In addition it should be real and positive.
In the exploratory calculation presented below we use \chem{KBr}, a material commonly 
used in infrared optics because of its transparency in that frequency range~\cite{KBr}. 
Its dielectric function varies little in the relevant frequency range, but the critical 
angle already depends significantly on frequency. The only condition for the metal 
layer is a large negative real and a nonvanishing imaginary part of the dielectric function 
for infrared frequencies. A common material choice for SPR is gold. We found a thickness 
of the gold film around 10\un{nm} to be optimal for our purpose. It is smaller than the 
50\un{nm} typically used in optical SPR~\cite{Kooyman08}.

The actual plasma wall of interest is the plasma-facing layer. Separated from the metal
by another dielectric layer, it is made out of the material whose charge accumulation
properties one wants to study. Since the dielectrics commonly used in plasma physics are 
electro-positive, and these are the ones we are aiming at, adding a separation layer 
with negative electron affinity confines the surplus electrons collected from the plasma 
to the plasma-facing layer. The separating (insulating) layer also prevents the surplus
electrons from spilling into the metal layer. Since in this work we focus on determining
the total amount of charge collected by the material in contact with the plasma, 
it is advantageous to make the plasma-facing layer rather thin. The insulating layer,
preventing the surplus electrons from leaving the film, leads then to a high local space 
charge density and thus to a high polarizability modifying the dielectric function of 
the film. It is this modification that makes the reflectivity of the stack charge-sensitive. 
We found a thickness of $d_2 = 20\,$nm to give satisfactory results. In our simulations
\chem{Al_2O_3} is used as the plasma-facing material, but other electro-positive dielectrics,
such as \chem{SiO_2}, could be used as well.The thickness of the insulating layer is not 
critical. We choose $d_1 = 40\un{nm}$, but even much thicker layers would not change the 
results significantly (see discussion below). For the material there are little restrictions. 
However, it is convenient if the infrared resonances of this layer are well separated from the 
resonances of the plasma-facing layer. We use \chem{MgO}. Since the densities of ions and electrons 
in the plasma are extremely low, the plasma is treated like a vacuum, that is, its dielectric
function $\varepsilon = 1$.

We investigate the reflectivity, that is, the ratio of the incident and reflected beam intensities 
as a function of the angle of incidence and the frequency of the impinging infrared light. As
usual, only p-polarized light is able to excite SPPs~\cite{Kooyman08}. Using the method of 
Lambin \textit{et al.} for multilayered materials~\cite{PRBLambin,PSLambin}, the solution of
the Maxwell equations yields an effective dielectric function $\xi_0(k,\omega)$ that can be
written as a continued fraction,

\begin{equation}
\label{eq:CF}
\xi_0 (k,\omega) = a_1 - \frac{b_1^2}{a_1 + a_2 - \frac{b_2^2}{a_2 + a_3 - \dots}}
\end{equation}
with

\begin{equation}
\label{eq:ai}
a_i = \frac{\varepsilon_i}{\sqrt{1 - \left(\frac{\omega}{kc}\right)^2\varepsilon_i} \,\tanh\left(\sqrt{1 - \left(\frac{\omega}{kc}\right)^2\varepsilon_i}~ k d_i\right) }
\end{equation}
and $b_i$ the same as $a_i$ when replacing $\tanh$ by $\sinh$.
Here $k = \omega/c\sqrt{\varepsilon_\mathrm{P}}\sin\alpha$ is the $y$-component of the wave vector
which is conserved throughout the system, $\varepsilon_i$ is the ($\omega$ dependent) dielectric 
function in layer $i$, $d_i$ is the layer's thickness and $c$ is the vacuum speed of light. For the
semi-infinite plasma layer the coefficients are $a_4 = 1/\sqrt{1-(\omega/(kc))^2}$ and $b_4 = 0$.
The value $\xi_0$ is the solution of a Ricatti equation at the prism-metal-interface, that is at 
$z=0$ (see fig.~\ref{fig.1}), and determines the reflectivity of the system via

\begin{equation}
\label{eq:R}
\abs{R}^2  = \abs{\frac{\xi_0 - \i \varepsilon_\ab{P} \tan\alpha}{\xi_0 + \i  \varepsilon_\ab{P} \tan \alpha}}^2~~.
\end{equation}
For the full derivation see ref.~\cite{PSLambin}, where the calculation is given without the prism,
but the adjustments to account for it are fairly simple.

In the infrared, the dielectric functions are highly frequency dependent.
Most dielectric materials can be modeled as a system of damped oscillators, so that the real and 
imaginary part of the dielectric function, labeled $\varepsilon'$ and $\varepsilon''$ respectively,
can be calculated as

\begin{equation}
\label{eq:eps'}
\varepsilon'(\omega) = \varepsilon_\infty + \sum\limits_{i}\frac{f_i \omega_i^2 (\omega_i^2 - \omega^2)}{(\omega_i^2 - \omega^2)^2 + \gamma_i^2\omega^2}
\end{equation}
and

\begin{equation}
\label{eq:eps''}
\varepsilon''(\omega) = \sum\limits_{i}\frac{f_i \omega_i^2 \gamma_i \omega}{(\omega_i^2 - \omega^2)^2 + \gamma_i^2\omega^2}~~.
\end{equation}
The values for the resonance frequencies $\omega_i$, the weighing factors $f_i$, the damping coefficients 
$\gamma_i$, and the limit values $\varepsilon_\infty$ are given in table \ref{tab.1} for \chem{MgO} and \chem{Al_2O_3}.
\begin{table}
	\caption{Material parameters for the dielectric functions of \chem{MgO}~\cite{MgO}, \chem{Al_2O_3}~\cite{Al2O3} and \chem{KBr}~\cite{KBr}. }
	\label{tab.1}
	\begin{center}
		\begin{tabular}{lccr}
			& \chem{MgO} & \chem{Al_2O_3} & \chem{KBr}\\
			\hline \hline
			$\varepsilon_\infty$ 		& 3.01 & 3.2	&1.39408	\\
			\hline
			$\omega_1 (\un{cm^{-1}})$ 	& 401	&385	&114.00\\
			$f_1$						& 6.6	& 0.3	&2.06217\\
			$\gamma_1 (\un{cm^{-1}}) $	& 7.619	& 5.58	&0\\
			\hline
			$\omega_2 (\un{cm^{-1}})$ 	& 640	&442	&164.99\\
			$f_2$						& 0.045	& 2.7&0.17673\\
			$\gamma_2 (\un{cm^{-1}})$	& 102.4	&4.42 	&0\\
			\hline
			$\omega_3 (\un{cm^{-1}})$ 	& 		& 569	&53476\\
			$f_3$						& 		& 3.0	&0.15587\\
			$\gamma_3 (\un{cm^{-1}})$	& 		& 11.38	&0\\
			\hline
			$\omega_4 (\un{cm^{-1}})$ 	& 		& 635	&57803\\	
			$f_4$						& 		& 0.3	&0.01981\\	
			$\gamma_4 (\un{cm^{-1}})$	& 		& 12.7	&0\\
			\hline
			$\omega_5 (\un{cm^{-1}})$	&		&		&68493	\\
			$f_5$						& 		& 		&0.79221\\	
			$\gamma_5 (\un{cm^{-1}})$	& 		& 		&0\\
			\hline
			$m^*/m$						&		& 0.4
		\end{tabular}
	\end{center}
\end{table}
For the gold layer, values from ref.~\cite{gold} were used and when necessary interpolated. 
In the infrared the absolute value of both real and imaginary part are large ($>1000$), the real part
being negative, and show roughly a $\omega^{-2}$ proportionality. The dielectric function of
\chem{KBr} is given as a Sellmeier equation and converted to the form of
eq.~(\ref{eq:eps'}) for convenience, but no imaginary part is considered.

Analyzing eqs. (\ref{eq:CF}) - (\ref{eq:R}), it becomes clear that the reflectivity $|R|^2$ will be unity 
if the dielectric functions have no imaginary parts, because then $\xi_0$ is real as well. Although the 
dielectric function of gold has a significant imaginary part in the whole infrared range, it only partakes
 in the absorption process through the surface plasmon. If the plasmon dispersion relation is not met, 
 there is no absorption by the gold layer.
\begin{figure} 
	\centering
	\includegraphics{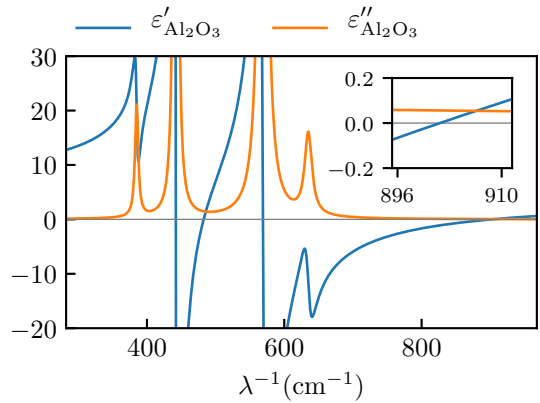}
	\caption{Bulk dielectric function of \chem{Al_2O_3} obtained from eqs.~\eqref{eq:eps'} 
        and~\eqref{eq:eps''} using 
        the parameters given in table~\ref{tab.1}. Absorption resonances occur when the imaginary 
        part of the dielectric function is large. The zero crossing of the real part at about 
        $\lambda^{-1} = 900\un{cm^{-1}}$ (see inset) will give rise to a Berreman resonance 
        in a film of thickness much smaller than the corresponding wavelength~\cite{Berreman}. 
	It is the mode we use for charge detection. Other resonances occur at lower wavenumbers, 
        but are significantly weaker and thus unsuitable for our purpose.}
	\label{fig.2}
\end{figure}
Taking the bulk dielectric function of \chem{Al_2O_3}--the material of interest we use as an
illustration--plotted in fig.~\ref{fig.2} into consideration, 
absorption frequencies can be identified. They are independent of the angle of incidence and occur 
where the imaginary part of the dielectric function is considerable compared to the real part, that is, 
at the resonance frequencies $\omega_i$, or where the real part crosses or approaches zero while the 
imaginary part stays finite, as it is the case near $\lambda^{-1} = 900\un{cm^{-1}}$. At this 
particular wavenumber, an enhanced absorption occurs for a film whose thickness is much smaller than 
the corresponding wavelength. In the infrared the film can be as thick as a few hundred 
nanometers for the resonance--which is called Berreman resonance \cite{Berreman}--to occur. It 
turns out to be very charge-sensitive and thus most suitable for our purpose because the 
additional polarizability in the film due to the surplus charges leads to a strong shift 
of the Berreman resonance. 

The polarizability $\alpha_\ab{P} = 4\pi \i\sigma_\ab{b} / \omega$, which is added to the dielectric 
function of the plasma-facing layer, is caused by the charges deposited into the plasma-facing 
layer. Using the memory function approach of ref.~\cite{Memory}, the bulk conductivity $\sigma_\ab{b}$ 
determining $\alpha_\ab{P}$ can be calculated as

\begin{equation}
\label{eq:sigma}
\sigma_\ab{b} (\omega) = \frac{e^2 n_\ab{b}}{m^*}\frac{\i}{\omega + M(\omega)}~~,
\end{equation}
where $e$ and $m^*$ are the electron charge and conduction band effective mass, and $n_\ab{b}$ is 
the bulk density of the surplus electrons. The memory function $M(\omega)$ takes electron-phonon
scattering into account via the interaction Hamiltonian $H_\ab{int} = \sum_{\vect{k},\vect{q}}
M c_{\vect{k} + \vect{q}}^\dagger c_{\vect{k}}(a_{\vect{q}} + a_{-\vect{q}}^\dagger)/(\sqrt{V} q)$,
with $M = \sqrt{2\pi e^2 \hbar \omega_\ab{LO}\left( \varepsilon_\infty^{-1} - \varepsilon_0^{-1}
	\right)}$, where $a_{\vect{q}}^{(\dagger)}$ and $c_{\vect{k}}^{(\dagger)}$ are the annihilation
 (creation) operators of phonons and electrons, respectively, and $V$ is the volume of the layer.
 To second order in $M$ the memory function is given by

\begin{equation}
\label{eq:M}
M(\omega) = M_0 \int\limits_{-\infty}^{\infty}\upd\bar{\nu} \frac{j(-\bar{\nu}) - j(\bar{\nu})}{\bar{\nu}(\bar{\nu} - \nu - \i 0^+)}
\end{equation}
with

\begin{equation}
\label{eq:j}
\begin{split}
j(\nu) = &\frac{\e^{\delta} }{\e^{\delta} -1}\abs{\nu + 1}\e^{-\delta(\nu + 1)/2}K_1(\delta\abs{\nu + 1}/2)\\
	&+\frac{1 }{\e^{\delta} -1}\abs{\nu - 1}\e^{-\delta(\nu - 1)/2}K_1(\delta\abs{\nu - 1}/2)~,
\end{split}
\end{equation}
where $\nu = \omega/\omega_\ab{LO}$ is the frequency in units of the longitudinal optical phonon frequency, 
$\delta = \hbar \omega_\ab{LO} /(k_\ab{B} T)$ is the LO phonon energy in units of the thermal energy (we use
 $T= 300\un{K}$), $K_1$ is a modified Bessel function, the prefactor in eq. (\ref{eq:M}) is $M_0 = 4e^2
 \sqrt{m^* \omega_\ab{LO} \delta}(\varepsilon_\infty^{-1} - \varepsilon_0^{-1})/(3\sqrt{(2\pi\hbar)^3})$,
  $\varepsilon_0$ is the static dielectric function, and $\omega_\ab{LO} = 807\, \un{cm^{-1}}$ is a longitudinal
   optical phonon frequency~\cite{Memory}.

\section{Results}
\begin{figure}
	\centering
	\includegraphics{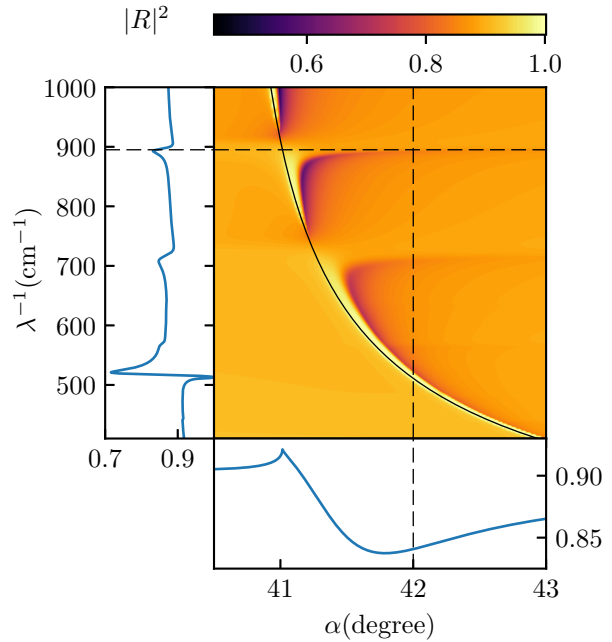}
	\caption{Reflectivity of the uncharged stack as a function of the angle of incidence $\alpha$ 
and the wavenumber $\lambda^{-1}$. The parameters are $d_\ab{M} =10\un{nm}$, 
$d_1 =40\un{nm}$ and $d_2 =20\un{nm}$. On the left the reflectivity is shown as a function 
of $\lambda^{-1}$ for a fixed angle of incidence $\alpha = 42^\circ$, indicated by the vertical 
dashed line, and on the bottom the reflectivity is plotted as a function of the angle of incidence 
for $\lambda^{-1} = 895\un{ cm^{-1}}$, indicated by the horizontal dashed line. The solid black
line gives the critical angle for each wavenumber. The top horizontal branch at about
$900\un{ cm^{-1}}$ is caused by the Berreman resonance of the \chem{Al_2O_3} layer, the lower one near
$700\un{cm^{-1}}$ is the Berreman mode of the \chem{MgO} layer, and the strong feature slightly 
above $500\un{cm^{-1}}$ is an ordinary SPR.}
	\label{fig.3}
\end{figure}
The reflectivity of the stack of materials without surplus charges is shown in 
fig.~\ref{fig.3} as a function of the wavenumber $\lambda^{-1}$ and the angle of incidence 
$\alpha$. Since the dispersion of SPPs is below the one of regular light, SPR occurs in our setup
only for angles larger than the critical angle $\alpha_\ab{c} = \arcsin\left(1/\sqrt{\varepsilon_\ab{P}}\right)$
which is wavenumber dependent because of the wavenumber dependence of the prism's dielectric 
function (solid black line). The SPP dispersion is the relation between the wavenumber and the angle
of incidence where absorption is observed. Because of the wavenumber dependence of 
$\alpha_\ab{c} $ the dispersion is bent 
over to larger angles. When another absorption mechanism occurs at the same wavenumber, like the Berreman 
resonance, avoided resonance crossing deforms the dispersion further, as can be seen for $\lambda^{-1}$ 
around $900\un{ cm^{-1}}$ and $700\un{ cm^{-1}}$. Far away form the critical angle, that is, far away 
from the black solid line, only the bulk absorption of the dielectric layers at these wavenumbers is 
observable and there is no angle dependence. However, approaching the critical angle, the horizontal 
absorption lines merge into the plasmon mode. Because the dispersion can be rather flat, when measuring 
the reflectivity as a function of the angle of incidence around these wavenumbers, a very broad 
minimum is observed compared to the narrow minimum resulting from the undisturbed plasmon dispersion.
This broad minimum in the angle of incidence shown in the bottom panel of fig~\ref{fig.3} for 
$\lambda^{-1}$ around $900\un{ cm^{-1}}$ is very sensitive to the wavenumbers. It will 
thus be modified when surplus charges change the dielectric function of the plasma-facing layer
and hence the zero-crossing of its real part. 
\begin{figure*}
        \begin{center}

                \includegraphics{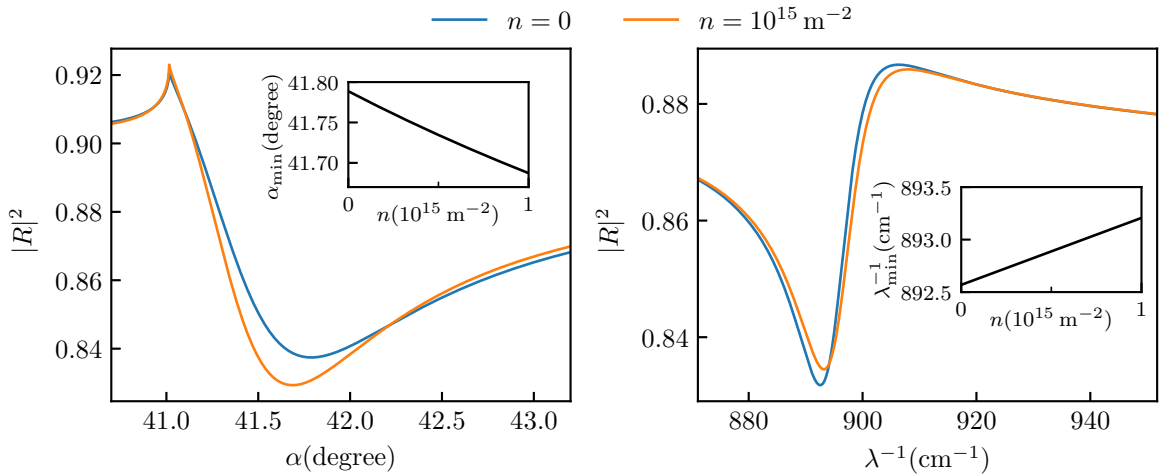}
                \caption{Reflectivity for different surface charges. On the left it is shown
as a function of the angle of incidence and $\lambda^{-1} = 895\un{ cm^{-1}}$ while on the right
it is plotted as a function of $\lambda^{-1}$ and $\alpha = 42^\circ$. The insets show the minimum
value of the dip as a function of the surface charge $n$, which is homogeneously distributed in the
plasma-facing layer giving rise to a space charge density $n_\ab{b}=n/d_2$. The parameters are the same
as in fig.~\ref{fig.3}. The shifts at the maximum density $n=10^{15}\un{m^{-2}}$ are $0.1014\,^\circ$
and $0.6377\un{ cm^{-1}}$.
                }
                \label{fig.4}
        \end{center}
\end{figure*}

Two practical ways are thus possible to measure a reflectivity curve in this type of 
setup. Either the wavenumber $\lambda^{-1}$ of the incident laser is fixed 
and the reflectivity is measured as a function of the angle of incidence $\alpha$, or the 
latter is fixed and the laser's wavenumber is varied. When surplus charges are added 
to the plasma-facing layer, the dispersion slightly changes because of the modification of the 
layer's dielectric function by the polarizability of the charges, and the dips in both measurement 
methods shift. As can be seen in fig.~\ref{fig.4} typical values for these shifts are $0.1^\circ$ 
in the angle and $0.6\un{cm^{-1}}$ in the wavenumber--or $8\un{nm}$ in the 
wavelength--for a surface charge density of $n = 10^{15}\un{m^{-2}}$ which is a rough estimate 
of the charge density to be expected based on the charge of dust particles in a low-temperature 
neon discharge~\cite{KRZ05}. These shifts should be measurable in common ATR setups which in the 
visible range achieve resolutions of about $10^{-3}$ degree or $0.1\un{nm}$. Refined setups
provide even resolutions up to $10^{-5}$ degree or $5\times10^{-4}\un{nm}$~\cite{TBH99,Jory95}. 
From the measured shift we can then determine the surface charge $n$ which for homogeneously 
distributed space charges obeys $n=n_\ab{b} d_2$ with $n_\ab{b}$ the bulk density and $d_2$ the 
thickness of the plasma-facing layer. In the inset of fig.~\ref{fig.4} we show how the 
minimum of the dips shifts as a function of the charge density. Measuring the position 
of the dip minimum opens thus a way to determine the surface charge $n$.

The shifts can be explained as follows: Considering that the additional charges shift the dielectric 
function linearly, and that the absorption mode occurs where the dielectric function crosses zero, it is 
quite clear that the surface charges will shift the dispersion upward in the vicinity of the Berreman 
mode. At a fixed wavenumber, the absorption dip as a function of the incident angle will thus 
shift to a lower angle, while for a fixed angle the dip will move to a higher wavenumber by 
about as much as the zero crossing of the dielectric function is shifted. 
To maximize the shift of the minimum angle, the dispersion should be as flat as possible at the 
chosen wavenumber. On the other hand, since the absorption becomes weaker as  
$\lambda^{-1}$ approaches the Berreman resonance, due to the avoided resonance crossing, the depth 
of the absorption dip is significantly reduced.
Thus, one needs to balance between sensitivity and absorption strength when choosing the parameters.
The data for the reflectivity dips and the charge-induced shifts of the reflection minimum shown in 
fig.~\ref{fig.4} were obtained for a particular choice of parameters. However, especially the angular 
sensitivity can be significantly enhanced by other choices of parameters, as we will now
discuss, but at the cost of flatter and broader absorption curves, that is, a decreased
detectability. 

In the rest of this section we describe the influence of the system parameters on the dispersion 
and the reflectivity curves shown in figs.~\ref{fig.3} and~\ref{fig.4}.
As mentioned above, the metallic layer is necessary for SPR, that is, for exciting SPPs. 
In the visible frequency range the optimal thickness $d_{\rm M}$  of the gold layer is 
about 50\un{nm}. It is imposed by two effects. Too thick layers reduce SPP excitation by too much 
absorption in the metal, while too thin layers lead to too high radiation damping in the prism 
attenuating thereby also the SPR. In our case, the SPR creates a weak angle dependence of the 
reflectivity near the Berreman resonance, which is in the infrared. To be of any use as a charge 
diagnostics it has to be detectable. We have thus to ensure that the metal layer is not too thick 
for most of the infrared radiation to be reflected at the prism-metal interface. Absorption by the 
SPPs or the modes of the dielectric bulk would then be too weak to produce a sizeable reflectivity
dip. For a thickness of $d_\mathrm{M} = 10\un{nm}$ we find about a 5 to 10\,\% drop at the minimum (see 
figs. \ref{fig.3} and \ref{fig.4}). If the layer is twice that thick the drop is only around 1 to 2\,\%.

The insulating dielectric layer underneath the metal is not involved in the absorption process 
at the relevant wavenumbers, 
because the Berreman resonance affiliated with this material is at a lower wavenumber, see 
fig.~\ref{fig.3}. Moreover, at the considered wavenumbers and angles the electromagnetic wave 
propagates through the insulating layer. Thus, its thickness $d_1$ is more or less arbitrary. Even 
for $d_1 > 1\un{\mu m}$ the shifts of the Berreman mode of the 
plasma-facing layer are still present. Only the avoided resonance crossing of the Berreman mode
of the insulating layer is somewhat suppressed. The particular numerical values of the shifts of 
the reflectivity dips, both in the angle of incidence and the wavenumber, vary with the thickness. 
For instance, for $d_1 = 1\un{\mu m}$ with the rest of the parameters as in fig. \ref{fig.3},
the shifts for $n = 10^{15}\un{m^{-2}}$ are $0.142^\circ$ and $0.386\un{cm^{-1}}$, while for 
$d_1 = 4\un{\mu m}$ the shifts are $0.076^\circ$ and $0.712\un{cm^{-1}}$.

The thickness $d_2$ of the plasma-facing layer has--in the present case, where we want to 
measure only the total amount of surplus charge, and hence use the layer also for charge 
confinement--a significant influence on the charge sensitivity of the method. It affects 
both the reflectivity dip in angle and in wavenumber. The reason is quite obvious since we 
assume the total surface charge $n$ provided by the plasma homogeneously distributed within 
that layer. Hence, the bulk charge density, entering the polarizability through the 
conductivity~\eqref{eq:sigma}, is given by $n_\ab{b} = n/d_2$. The thicker the plasma-facing 
layer the smaller is therefore $n_\ab{b}$ and hence the polarizability driving the shifts of the 
reflectivity minima. The larger $d_2$ the less pronounced is thus the reflectivity dip 
as a function of $\lambda^{-1}$ for a fixed angle making it thus less suitable for
charge diagnostics. However, in the setup we use a thicker layer implies also that the 
avoided resonance crossing becomes stronger, that is, the flat branch of the dispersion 
at around $900\un{ cm^{-1}}$ (viz: fig.~\ref{fig.3}) degrades already at larger angles.
As a result, the reflectivity dip as a function of the angle becomes wider and less deep. 
But surprisingly it shifts stronger with the surface charge density $n$ than the 
narrower dip of a less thick layer. Thus, by choosing the thickness $d_2$  
accordingly, the charge sensitivity of the reflectivity dip as a function of angle 
for fixed $\lambda^{-1}$ can be enhanced. Pushing the laser frequency closer 
to the Berreman resonance has the same effect. It makes the reflectivity dip flatter and 
wider but at the same time also more charge-sensitive.  

\section{Conclusion}

We showed that in an infrared ATR setup the presence of surplus charges deposited into a 
plasma-facing dielectric layer manifests itself in a shift of a reflectivity dip both 
in the wavenumber and the angle of incidence. The results we obtained suggest moreover 
that the shifts are detectable by standard infrared equipment. In this exploratory
work we focused on detecting the total charge accumulated in the plasma-facing film
which we moreover assumed to be homogeneously distributed. The thicknesses of the 
layers of the stack used as a charge measuring device could thus be chosen freely 
to optimize the dip's detectability and charge sensitivity. 
In principle the device can also be used to map out the density profile normal to 
the interface. The plasma-facing layer then has to be thick enough to host the whole
space charge profile. More refined theoretical treatments are then necessary. The 
principle of the method however remains the same: Using the Berreman mode of the plasma-facing
layer as a charge sensor. Compared to other approaches measuring the wall charge,
the method we suggest does not exploit material-specific properties. Being a spectroscopic
technique it may have the potential to track the charge accumulation in time. 
It does not require complex experimental setups. In fact 
we expect it to be compatible with commonly used discharge geometries. The stack of 
materials measuring the wall charge can be integrated into the plasma wall or 
the electrode. Mechanical stability is then provided by a sufficiently thick prism.

\bibliography{ref}

\end{document}